\documentclass{aa}
\usepackage[utf8]{inputenc}
\usepackage{amsmath}
\usepackage{amsfonts}
\usepackage{amssymb}
\usepackage{color,soul}
\usepackage{longtable}
\usepackage{graphicx, bm}
\usepackage{booktabs}
\usepackage{tablefootnote}
\usepackage{placeins}
\usepackage{subfig}
\usepackage[export]{adjustbox}

\begin{document}

\title{Varying water activity and momentum transfer on comet 67P/Churyumov-Gerasimenko from its non-gravitational forces and torques}
\titlerunning{Varying activity on comet 67P from its non-gravitational forces and torques}

\author{N.~Attree \inst{1, 2} \and P.~Gutiérrez \inst{1} \and O.~Groussin \inst{3} \and J.~Bürger \inst{2} \and H.~U.~Keller \inst{4} \and T.~Kramer \inst{5} \and R.~Lasagni Manghi \inst{6} \and M.~Läuter \inst{7} \and  P.~Lemos \inst{2} \and J.~Markkanen \inst{2} \and R.~Marschall \inst{8} \and C.~Schuckart \inst{2}}

\institute{Instituto de Astrofísica de Andalucía - CSIC,
Glorieta de la Astronomía s/n, 18008 Granada, Spain
(\email{attree@iaa.es})
\and
Institut f{\"u}r Geophysik und Extraterrestrische Physik, Technische Universit{\"a}t Braunschweig, Mendelssohnstr. 3, 38106 Braunschweig, Germany
\and
Aix Marseille Univ, CNRS, CNES, Laboratoire d'Astrophysique de Marseille, Marseille, France
\and
DLR Institut für Planetenforschung, Rutherfordstraße 2, 12489 Berlin, Germany
\and
Institute for Theoretical Physics, Johannes Kepler University Linz, Austria
\and
Alma Mater Studiorum - Universit{\`a} di Bologna, Dipartimento di Ingegneria Industriale, Via Fontanelle 40, I-47121 Forl{\`i}, Italy
\and
Zuse Institute Berlin, 14195 Berlin, Germany
\and
CNRS, Laboratoire J.-L. Lagrange, Observatoire de la Côte d’Azur, Boulevard de l’Observatoire,
CS 34229 - F 06304 NICE Cedex 4, France
}

\abstract{}
{We investigate the ability of a simultaneous fitting of comet 67P/Churyumov-Gerasimenko's non-gravitational forces, torques and total water-outgassing rate, as observed by Rosetta, to constrain complex thermophysical models of cometary material.}
{We extend the previous work of fitting geographically defined surface outgassing models to the Rosetta observations by testing the effects of a more detailed geomorphological mapping, the resolution of the shape-model used, self-heating by neighbouring facets on the shape-model, thermal inertia in the outgassing solution, and variation in the momentum coupling between the gas and the nucleus. We also directly compare the non-gravitational acceleration curves available in the literature.}
{We correct an error in the calculation of pole-orientation in the previous paper. We find that, under the assumptions of the model: non-gravitational forces and torques are driven by water sublimation from the nucleus, thermal inertia and self-heating have only minor effects, spatially uniform activity cannot explain 67P's non-gravitational dynamics, spatially uniform momentum transfer cannot explain 67P's non-gravitational dynamics, and different terrain types have different instantaneous responses to insolation.}
{Consolidated terrain facing south on 67P/Churyumov-Gerasimenko has a high outgassing flux, steep response to insolation, and large gas momentum transfer coefficient. Meanwhile, that facing north behaves differently, producing low-to-no water outgassing, and with a lower momentum transfer efficiency. Dusty terrain also has a lower outgassing rate and momentum transfer efficiency, and either depletes its volatile component or is buried in fall-back as the comet approaches the Sun. Momentum transfer appears correlated with insolation, likely due to an increased enhancement in the gas temperature as the dust it flows through is heated.}
   

\maketitle
 
\section{Introduction}

Cometary activity, the sublimation of volatile ices and the ensuing ejection of non-volatile dust particles, is controlled by the composition and structure of the surface material, which is potentially amongst the most primitive accessible in the solar system. Therefore, the study of outgassing rates over the surface of a particular comet, alongside the resulting non-gravitational force and torque on the comet's nucleus, provides important insights into the era of planet formation, as well as the subsequent evolution of the surface. For comet 67P/Churyumov-Gerasimenko (hereafter 67P), the investigation of ESA's Rosetta spacecraft over a period of two years has resulted in large amounts of data being gathered, but no unified model of cometary activity can explain all the observations. It is therefore worth re-examining the available data together in order to place further constraints on thermophysical models of cometary material.

The back-reaction force generated by the outgassing produces a net acceleration (non-gravitational acceleration; NGA) that can be detected in 67P's trajectory from the ground \citep[e.g.][]{Davidsson05, Gutierrez05} and by Rosetta \citep{Mottola2020}; as well as a non-gravitational torque (NGT) that changes its spin-period \citep{Keller} and pole-axis orientation. Previous works (e.g.~\citealp{Attree2019, Kramer2019, kramer2019b}) have attempted to fit these observed changes, alongside the total water outgassing rate, to simple models of surface activity, but no single model was able to explain all the data. Likewise, more complex thermophysical models, such as that of \citet{Davidsson2022} and \citet{Fulle2020}, the latter as applied by \citet{Attree2023}, also struggle to reproduce the NGA curves without post-hoc optimisation of the parameters. 

As a brief description of the procedure involved in our study, in our previous paper \citep{Attree2023}, a simple surface sublimation model was parameterised in terms of the geographic regions defined on 67P's surface \citep{Thomas2018} in order to try and match the NGA extracted by \citet{Bologna}; the NGT; and the total water-production rate of \citet{Hansen}. In our procedure, pole-orientation evolution is intentionally left out of the fitting process but it is used to verify the quality of our fit by comparing the modelled change derived from the solution with its actual observational change. The fit showed an improvement over previous attempts, but still struggled to reproduce the non-radial components of the NGA simultaneously with the NGT. Additionally, several factors were not considered in detail such as the resolution of the shape-model used, the effects of surface self-heating and thermal inertia, and potential variability in the momentum coupling between the gas and the nucleus. Additional datasets for the NGA \citep{kramer2019b, Farnocchia}, water-production rate \citep{Laeuter2020}, and surface geomorphological mapping \citep{Birch17} are also available but were not previously used. We therefore extend the previous work by incorporating the above factors, while seeking to a find a still relatively simple model for parameterising surface activity that can fit the observations and be used as a constraint or target for the output of more complex thermophysical models.

The rest of the paper is organised as follows: we first review and compare the various extractions of 67P's non-gravitational accelerations from its trajectory in Section \ref{method}. Next, in Sections \ref{check} and \ref{results} we examine the effects of shape-model resolution, self-heating, thermal inertia, and varying the momentum transfer coefficient on the modelling of 67P's NGAs and NGTs. In particular, in Section \ref{results:bestfit}, we present our best-fitting surface activity model. We discuss the implications of these results for models of cometary material in Section \ref{discussion}, and conclude in Section \ref{conclusion}.

\section{67P's non-gravitational acceleration}
\label{method}

\begin{figure*}
\begin{tabular}[b]{@{}cc@{}} \\[-\dp\strutbox]
\subfloat{\includegraphics[width=6cm]{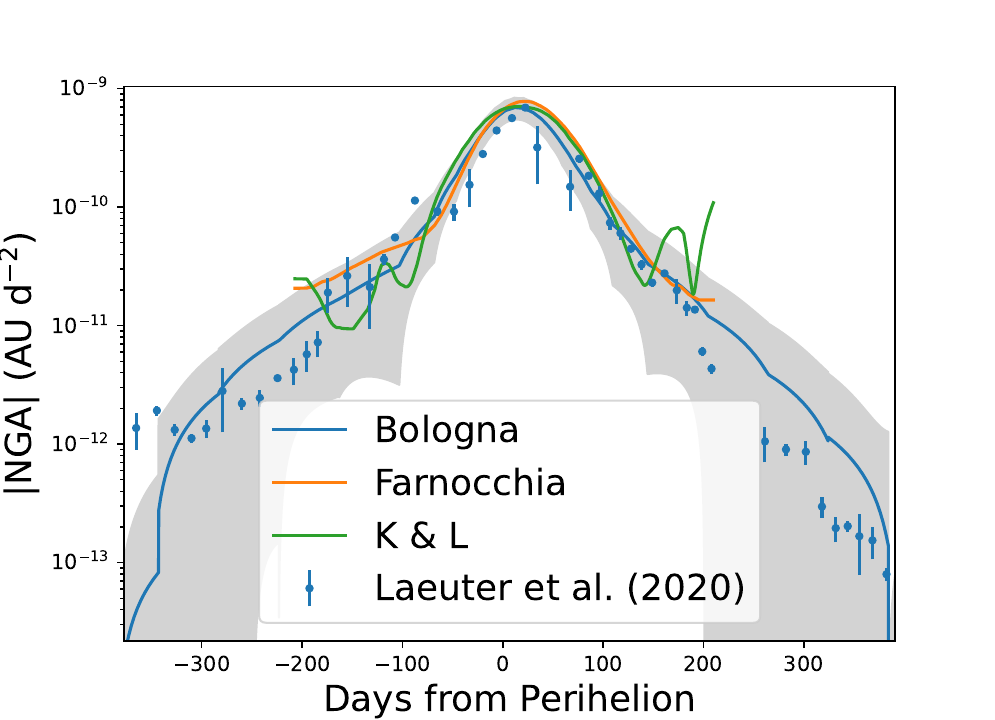}}
\subfloat{\includegraphics[width=6cm]{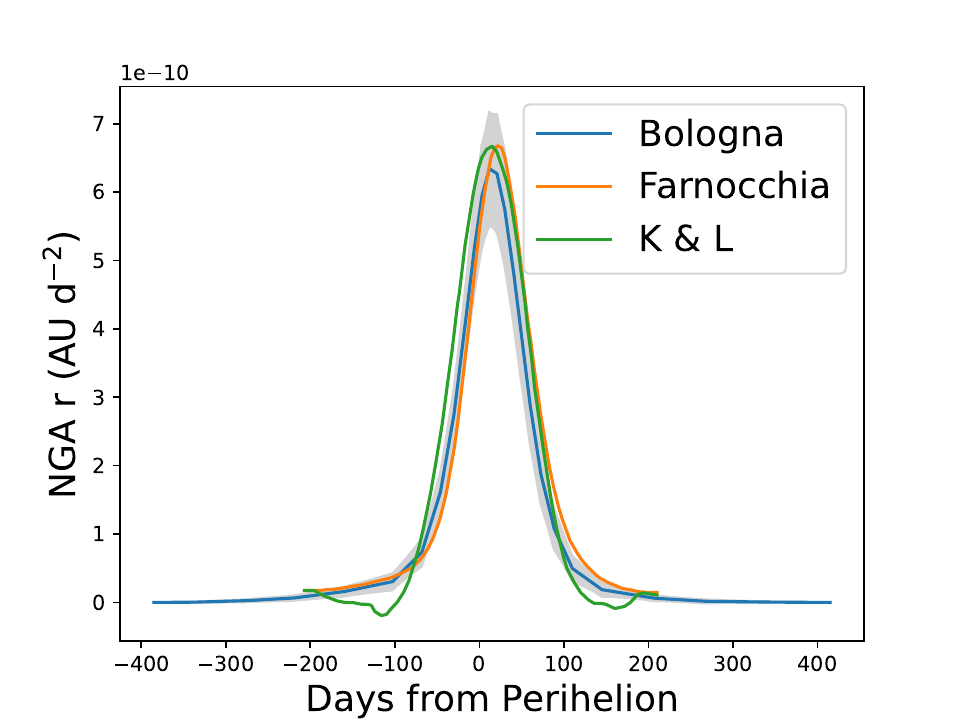}}
\\
\subfloat{\includegraphics[width=6cm]{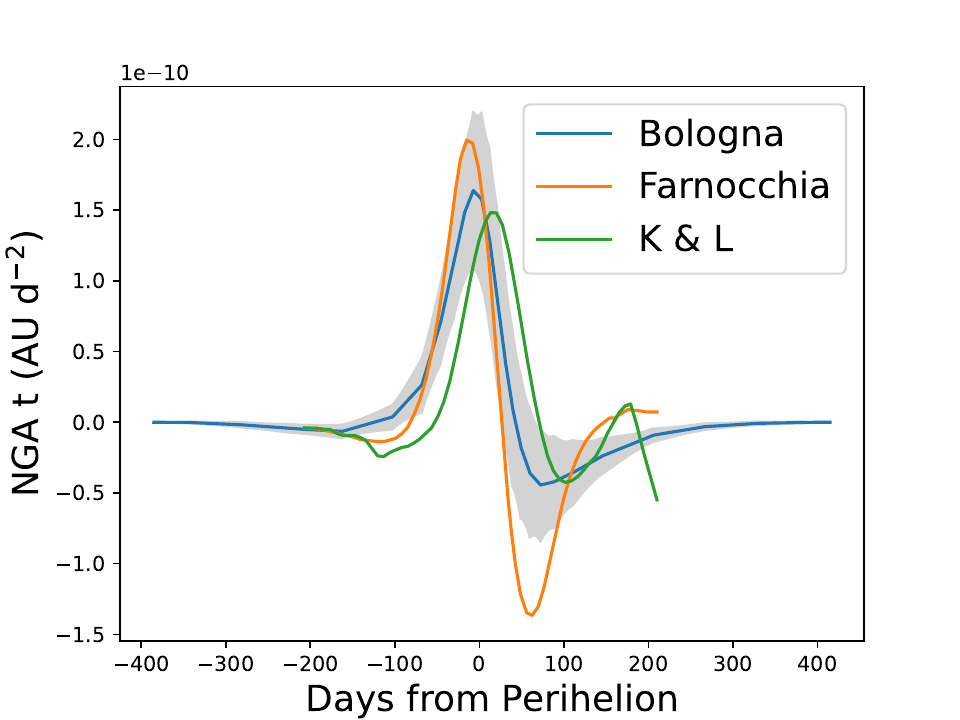}}
\subfloat{\includegraphics[width=6cm]{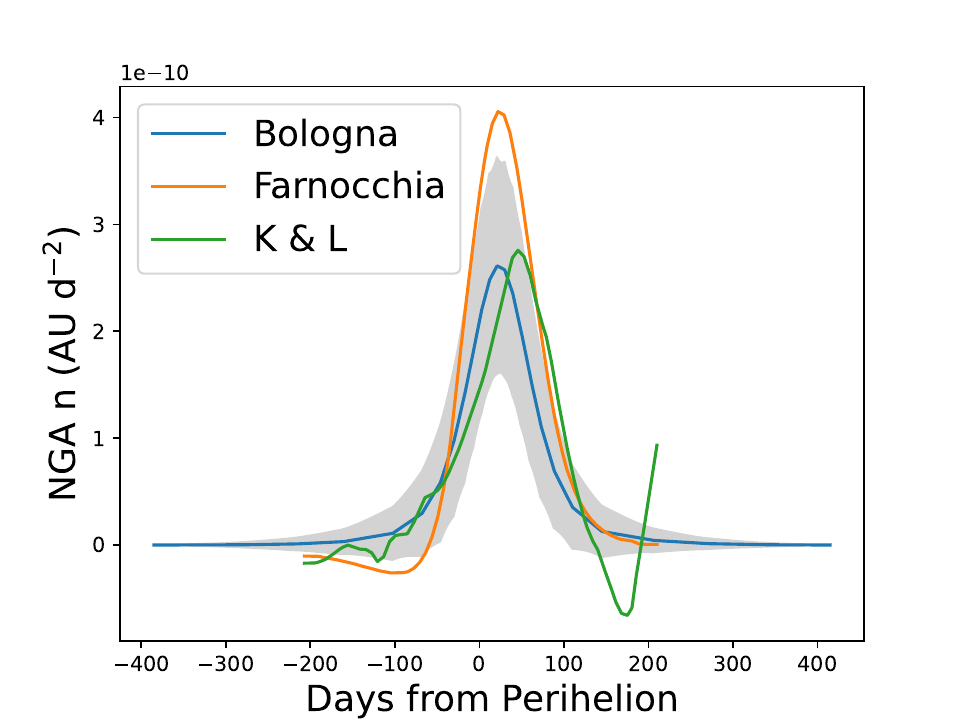}}
\end{tabular}\hfill
\begin{minipage}[b]{0.3\textwidth}
\caption{Extracted 67P non-gravitational acceleration curves, in total magnitude and the three components of the cometo-centric RTN frame, from \citet{Bologna} (labelled Bologna), \citet{Farnocchia}; and \citet{kramer2019b}. Grey bands show the formal 10-sigma uncertainty estimates from \citet{Bologna}, while the other models' uncertainties are omitted for clarity. The magnitude plot also shows the total water-production rate of \citet{Laeuter2020} scaled as described in the text.}
\label{Plot_NGAcomparison}
\end{minipage}
\end{figure*}

Before optimising the NGA models, we first re-examine the data that they will be assessed against. The non-gravitational torques have been described elsewhere \citep{Attree2019, Kramer2019}, but we below present a brief comparison of the accelerations extracted by three different authors: \citet{Bologna, Farnocchia}; and \citet{kramer2019b}. As described in the references, \citet{Farnocchia} fit a rotating-jet model to a limited set of high-accuracy Rosetta radio-ranging data and ground-based astrometry collected before and after the perihelion, while \citet{kramer2019b} fit smoothed curves to the residuals of the available SPICE kernels from the ESA flight dynamics solution. Conversely, \citet{Bologna} uses a stochastic acceleration model and processes the full set of Rosetta radiometric data (ranging and differential one-way ranging: $\Delta$DOR) and ground-based astrometry available on the Minor Planet Center. To account for degraded accuracy and possible measurement biases close to perihelion, a weighting scheme based on the spacecraft-to-comet distance has been applied. The \citet{Bologna} accelerations have been updated since being first presented in \citet{2021EGU} and used in \citet{Attree2023}, and we therefore compare the three curves here, with their three components in the cometo-centric RTN frame plotted, along with the total magnitude, in Figure \ref{Plot_NGAcomparison}. Also plotted for comparison is the total water-production rate, $Q_{H_{2}O}$ from \citet{Laeuter2020}, scaled to the NGA magnitude by the equation of e.g.~\citet{Jewitt_2020}, 

\begin{equation}
\mathrm{NGA}=\frac{k_{R} v_{av} Q_{H_{2}O}}{M_{67P}}.
\label{NGF}
\end{equation}
Here the orbitally averaged momentum transfer coefficient for 67P is estimated to be $k_{R}\approx0.5$ by \citet{Jewitt_2020}, and we use an orbitally averaged gas-velocity of $v_{av}=500$ ms$^{-1}$ (a thermal velocity corresponding to around 200 K), along with the known comet mass $M_{67P}$ \citep{Patzold16}.

Using these reasonable values, the NGA magnitude scales well with the heliocentric evolution of the water-production rate, suggesting that most of the NGA is produced by water sublimating from the cometary surface (extended sources of water in the coma would not contribute a back-reaction force on the nucleus). The agreement between NGA magnitude and  water-production rate also supports the idea that NGAs from other gas species are negligible \citep{Davidsson2022, Attree2023}, with the possible exception of CO$_{2}$, which exceeds water production after the second equinox. The \citet{Bologna} NGA curve, although small in overall magnitude here, slightly exceeds the scaled water production from 200 days after perihelion, possibly hinting at a CO$_{2}$ outgassing contribution. Nonetheless, we concentrate for the remainder of this paper on H$_{2}$O emission which dominates the large magnitude accelerations found around perihelion.

The total magnitudes of the three NGA extractions are very similar. In fact, the NGA vector generally agrees across the three works, with some minor differences. The curves of \citet{kramer2019b}, for example, contain more noise than the smooth rotating-jet model of \citet{Farnocchia}, while also peaking slightly later after perihelion in the transverse and normal components. The rotating-jet model has slightly more acceleration in these two components than the other two models, although its overall magnitude is still similar. Overall though, there is good agreement between the extractions, while the uncertainty bounds on the \citet{Bologna} fit seem well estimated as they generally encompass the other curves. We therefore feel confident in focusing on the \citet{Bologna} acceleration and uncertainty estimates in the following sections, although we have also performed test-optimisations of the activity model (see below) to all the available acceleration curves without drastic changes in conclusion.

\section{Resolution, self-heating, and thermal inertia}
\label{check}

In an effort to improve the match of previous models to the available data, we now investigate some background aspects in the modelling that have, until now, mostly been assumed. These are the effects of shape-model spatial resolution, facet self-heating, and thermal inertia.

\begin{figure}
\subfloat{\includegraphics[scale=0.5]{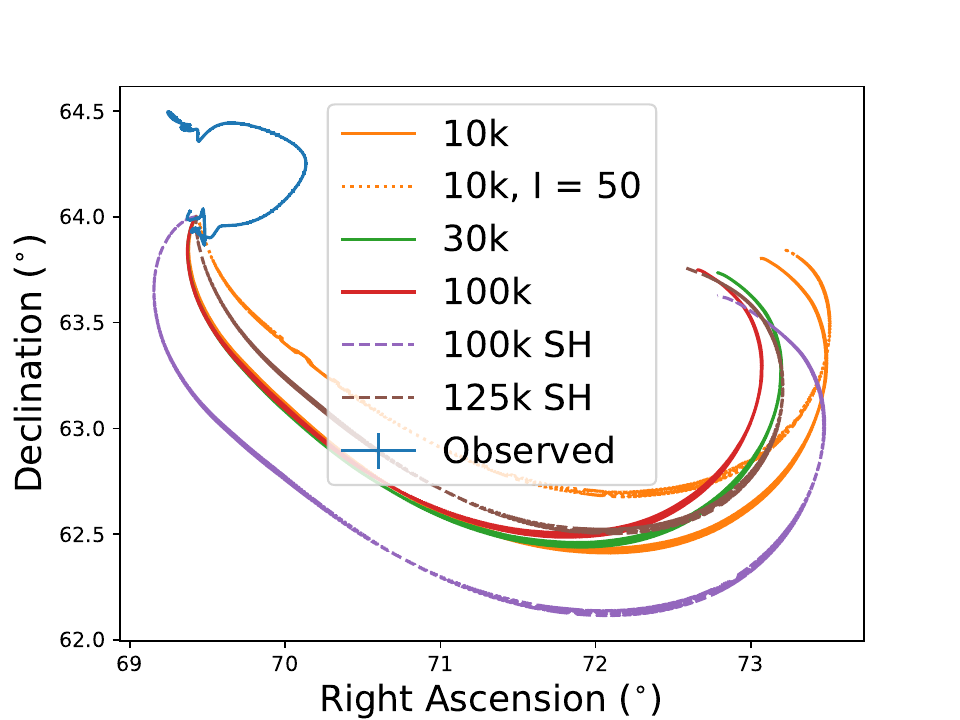}}%
\qquad
\subfloat{\includegraphics[scale=0.5]{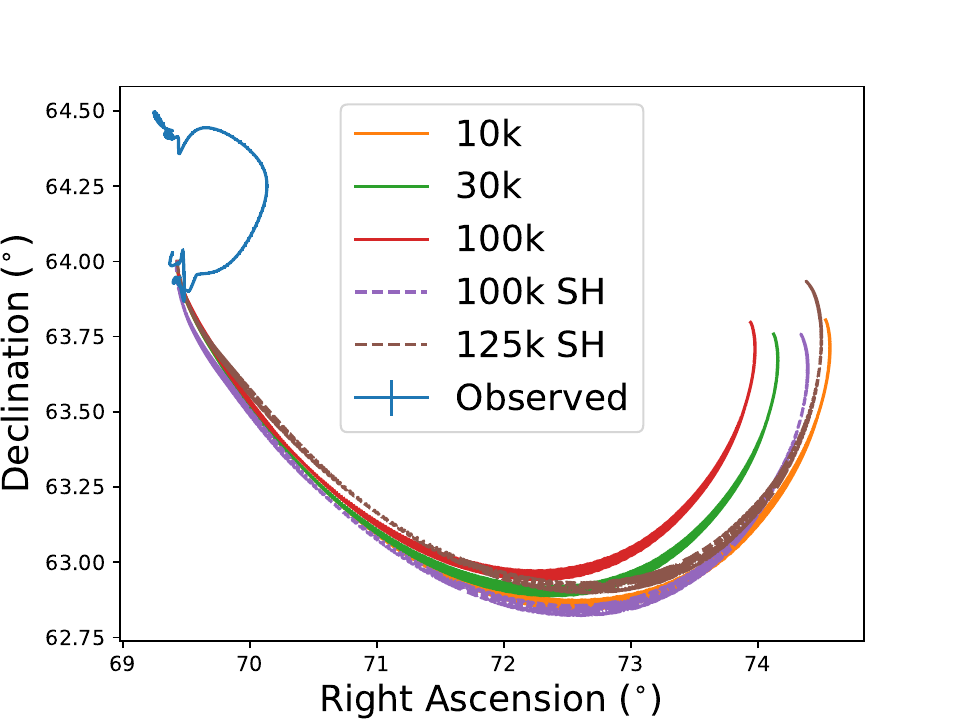}}%
\caption{Observed pole orientation evolution (RA/Dec) as compared to results for different shape-model resolutions, with and without self-heating (SH) and thermal inertia. Top: for a uniform EAF of $8\%$, bottom: for the model of \citet{Fulle2020}. The observations have been smoothed and error bars are plotted but are small at this scale.}
\label{Plot_resultsRAdec_check}
\end{figure}

Resolution was discussed in \citet{Attree2019} and \citet{Kramer2019}, but no comprehensive analysis of the effects of different shape-model resolutions on the NGAs and NGTs has been carried out. In an effort to find any systematic trends, we therefore ran the exact same analysis method with four different versions of the same shape-model (SHAP7; \citealp{Preusker17}) with resolutions: 10, 30, 100, and 125 thousand facets (the latter being the same as used in \citealp{Attree2019, Attree2023}). Identical water-production rate curves were found, while NGAs and, in particular NGTs were seen to vary slightly. In Figure \ref{Plot_resultsRAdec_check}, we show the resulting pole-orientation evolutions, following the algorithm of \citet{Attree2023} and \citet{Kramer2019}, for models with uniform activity and an effective active fraction (EAF) of $8\%$ (top), and the \citet{Fulle2020} model (bottom). We focus here on the pole-orientation because it is very sensitive to the shape of the nucleus. Nonetheless, no significant trends are observed beyond some scatter in the results.

Next, we turned on self-heating by neighbouring facets on the 100k shape model, using the algorithm described in detail in \citealp{Attree2019}, for comparison with the results of that paper. Some changes are seen in the pole evolution, particularly for the $8\%$ EAF model; while the results for the \citet{Fulle2020} model, which produces a much closer fit to the observed water production, are tightly clustered.

Finally in this analysis, we included thermal inertia by using a different thermal model \citep{Groussin2013} that solves the 1D heat transport equation for each facet for a simple bulk material with a thermal inertia equal to 50~Jm$^{-2}$s$^{-0.5}$K$^{-1}$, typical of 67P \citep{Gulkis2015}. Due to the long computing times to generate a heat conductivity map over a complete orbital revolution, we used the shape model with 10 thousand facets. As before, this was run once with no sublimation, to simulate a pure-dust surface, and once with sublimation at the surface, to simulate pure-ice, and the results scaled by a constant $8\%$ EAF. This thermal model is not compatible with our implementation of the \citet{Fulle2020} model. Figure \ref{Plot_resultsRAdec_check} (top) shows that the effects of thermal inertia on the pole evolution of the 10k model are limited, with only a minor rotation of the trajectory of the pole-orientation between the cases with and without thermal inertia. We are therefore confident that spatial resolution and the presence or absence of self-heating and a limited thermal inertia do not have a large effect on the pole-orientation evolution.

During the course of this investigation, an error in the code used for computing the pole-orientation evolution in \citet{Attree2023} was discovered. The error has been corrected and the pole-orientation evolution of the best-fitting solutions from that paper is recomputed below.

\begin{figure}
\subfloat{\includegraphics[scale=0.5]{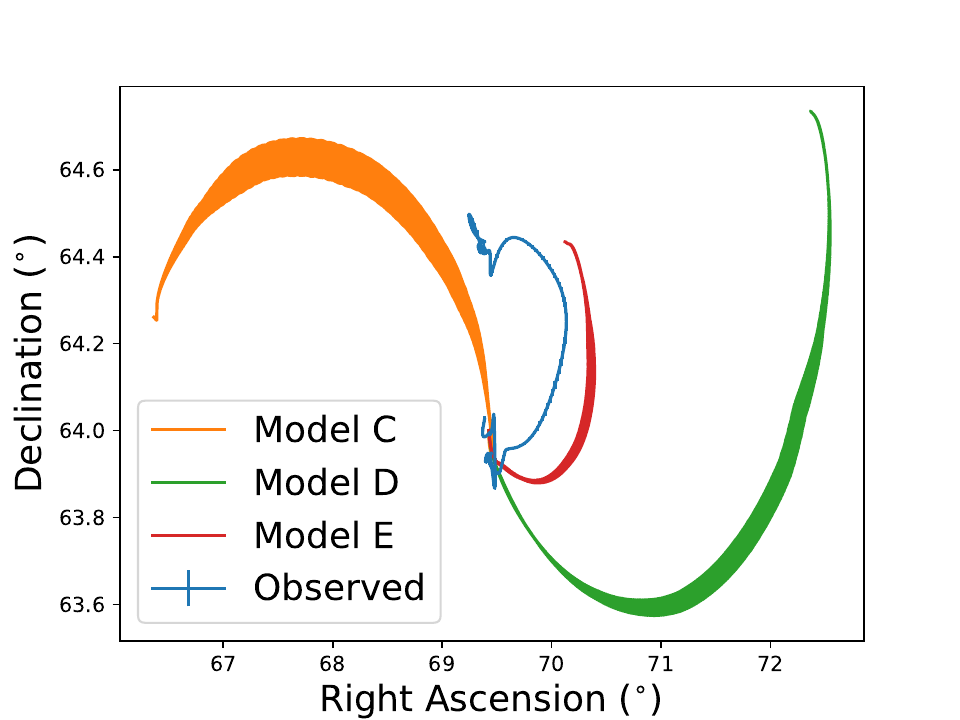}}%
\qquad
\subfloat{\includegraphics[scale=0.5]{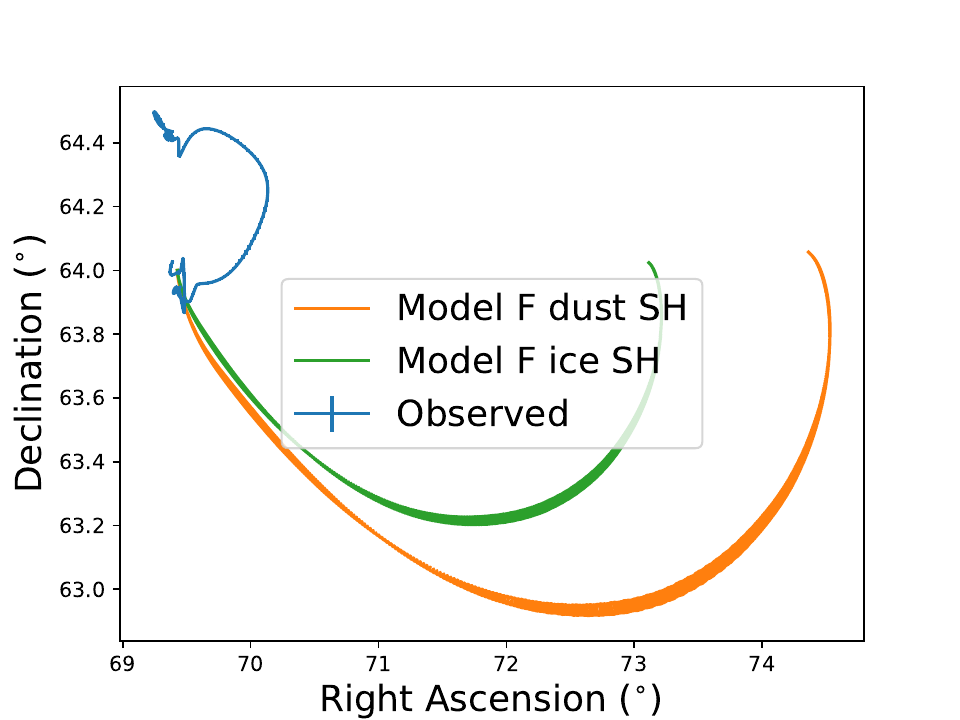}}%
\caption{Observed pole orientation evolution (RA/Dec) compared to the corrected solutions of models C, D, and E and the two versions of model F from \citet{Attree2023}. The thickness of the model lines is due to the daily oscillations. Compare to Figures 9 and B4 in \citet{Attree2023}.}
\label{Plot_resultsRAdec_corrected}
\end{figure}

Figure \ref{Plot_resultsRAdec_corrected} shows the results for three different splits of the surface (labelled C, D, and E; see \citealp{Attree2023} for details) with a surface energy-balance thermal model and spatially-varying and time-varying effective active fractions (EAF). It can been seen that the pole-orientation evolution is qualitatively the same as in \citet{Attree2023}, Figure 9, with only minor differences in the values. Given the inherent modelling uncertainties discussed above, the previous conclusions regarding these best-fit models are unchanged. On the other hand, the implementation of the different activity model of \citet{Fulle2020} as model F does now show a qualitatively different behaviour, with the pole orientation following a different trajectory. Comparing Figures \ref{Plot_resultsRAdec_check} and \ref{Plot_resultsRAdec_corrected}, we see that a spatially uniform activity pattern will always result in the pole orientation moving in the same direction. This result can be checked by comparison with the uniform activity model of \citet{Kramer2019} (green curve, Fig. 6), reassuring us that the updated code is correct.

To conclude this preliminary work, we see that: variations in shape-model resolution, and the presence or absence of self-heating and a moderate thermal inertia, have very little effect on the overall shape of the NGA and NGT curves. We also note that uniform activity distributions always produce roughly the same pole-orientation evolution, and that this evolution is not compatible with the observations. We may now move on to more complicated, spatially varying, activity models.

\section{Momentum coupling}
\label{results}

As described above, previous studies have struggled to simultaneously reproduce the observed non-radial components of the NGA with the NGTs at perihelion. We have seen that varying self-heating and thermal inertia with a uniform surface activity has little effect on the pole evolution; so, as in previous studies, we now consider activity models that divide the surface into super-regions, each with their own time-varying EAF. Further, we now investigate a variable momentum transfer coefficient, $\eta$, which was previously assumed to be constant over the comet's surface, being fixed at the beginning of each run. All optimisations in this section are performed with the algorithm described in the previous two papers \citep{Attree2019, Attree2023}, with data weights adjusted to give each dataset approximately equal contributions to the overall fit as before.

The momentum transfer coefficient, $\eta$, describes the collimation of the gas flow, relative to sublimation from a flat pure-ice surface, whilst also considering the amount of backwards, return flux as re-condensation from the Knudsen layer \citep{Crifo1987}. The amount of collimation and return-flux should, in principle, depend on the exact location of the gas sublimation and how it flows through the material, i.e.~at the surface or at depth, and between or within pebbles. Meanwhile, the total gas momentum is then the product of this factor with the mass flow-rate and the gas velocity, which is controlled by its temperature. We previously assumed this gas temperature as equal to the calculated surface temperature for a pure dust surface ($T_{dust}$ in Eqn.~5 of \citealp{Attree2019}), and assumed a fixed $\eta$ value corresponding to surface sublimation. Considering the variation in surface texture across the comet, and the limitations of the surface sublimation model, it is likely that both of these factors should vary. A proper treatment of this problem requires a full thermophysical model, but here we can consider a simple modification to make our description more realistic. We first split the effects described above into a material-dependent, and an insolation-dependent component. The first takes into account in a rough way the depth structure of the outgassing material, which likely varies across the comet, by simply parameterising a different momentum transfer coefficient for each super-region. The second part accounts for a potentially enhanced insolation dependence of the emitted gas-temperature compared to a simple surface-sublimation model. For example, \citet{Skorov2024} show that gas flowing through a porous dust-layer can acquire a temperature somewhere between its initial sublimation temperature, $T_{ice}$, and the dust temperature, with the roughly linear dependence on the insolation. We therefore rewrite our expression for the magnitude of the outgassing force per facet as
\begin{equation}
F = \eta \cdot \mathrm{EAF} \cdot Z \sqrt{\frac{8R_{gas}T_{ice}}{\pi M}} \cdot \left( \frac{T_{dust}}{T_{ice}}\right),
\label{eta}
\end{equation}
where $\eta$ and EAF are the local material momentum transfer coefficient and a time-dependent effective active fraction (given by Eqn.~\ref{gauss} below), $Z$ is the local sublimation rate determined by the energy balance, and $R_{gas}$ and $M$ are the gas constant and molecular mass. The gas and dust temperatures are solved for pure water-ice and pure dust surfaces as described in the previous papers, resulting in a factor in brackets ranging between 1 and $\sim1.9$ and correlated with increasing insolation (see Fig.~$\ref{Plot_insolation}$). We can consider the product of this factor with $\eta$ as the 'full' momentum transfer coefficient in our model, while bearing in mind that it is dependent on the specific energy-balance model used here and that comparisons with values derived using other thermophysical models are complicated.

We re-ran the best-fit solutions of \citet{Attree2023} (solutions D and E) with the updated momentum coupling model, while also modifying the way EAF levels are parameterised. Instead of the previous time-variation model that used a base and a peak EAF level, linked by half-Gaussian profiles, we now multiply the peak EAF by a Generalised Asymmetric Gaussian in time, e.g.~\citet{Gaussian}:
\begin{equation}
\mathrm{EAF (t)} = \left\{ \begin{array}{lcr}
  \mathrm{EAF_{peak}} \mathrm{exp}\left[ -\left(\frac{t_{0} - t}{t_{r}}\right)^{s}\right], \quad \text{for} \quad t < t_{0},\\
  \mathrm{EAF_{peak}} \mathrm{exp}\left[ -\left(\frac{t - t_{0}}{t_{f}}\right)^{s}\right], \quad \text{for} \quad t > t_{0},\\
\end{array} \right.
\label{gauss}
\end{equation}
where each super-region is then parameterised by its peak EAF level, peak time $t_{0}$, widths of its rise-time $t_{r}$, and fall-time $t_{f}$, and a shape factor that allows the profile to be adjusted to be more or less steep than the standard $s=2$ Gaussian. Alongside the spatially varying momentum  transfer coefficient, this leads to 6 free parameters per each of the 5 or 6 super-regions, for a total of 30 or 36 parameters (for comparison, \citealp{Kramer2019} also have 36 parameters).

Some small improvements to the fit were achieved, particularly in the R and T components of the NGA, as the asymmetric shapes and varying momentum transfer coefficients relaxed the limits on previous solutions. However, similar problems remained, as discussed previously, in terms of simultaneously fitting the NGA and NGT. In particular, the fact that the super-regions defined in \citet{Attree2023} models D and E cover areas of both cometary hemispheres limits their ability to generate the large non-radial NGA peaks seen at perihelion.

At this point we also attempted a fit using the per-facet geological mapping of \citet{Birch17}, instead of the rather broad geographic regions defined by \citet{Thomas2018}. The mapping of \citet{Birch17} divides the surface into seven terrain types, which broadly follow the previous classification into generally dusty and generally rocky terrains, with some more precise differentiation of different types of dusty plain and talus materials.

Unfortunately, neither fits to the full seven terrain types ($7\times6=42$ free-parameters), nor a grouping into `Dusty' and `Rocky' terrains ($2\times6=12$ free-parameters) could produce substantial improvements to the results above, or to the results of our previous work. Looking at the maps of Figure 7 in \citet{Birch17}, it can be seen that many terrain types extend over both comet hemispheres, in particular the `cliff' terrain that dominates activity at perihelion. We thus return to the same problem discussed above.

\section{Best-fitting surface models}
\label{results:bestfit}

\begin{figure}
\subfloat{\includegraphics[scale=0.5]{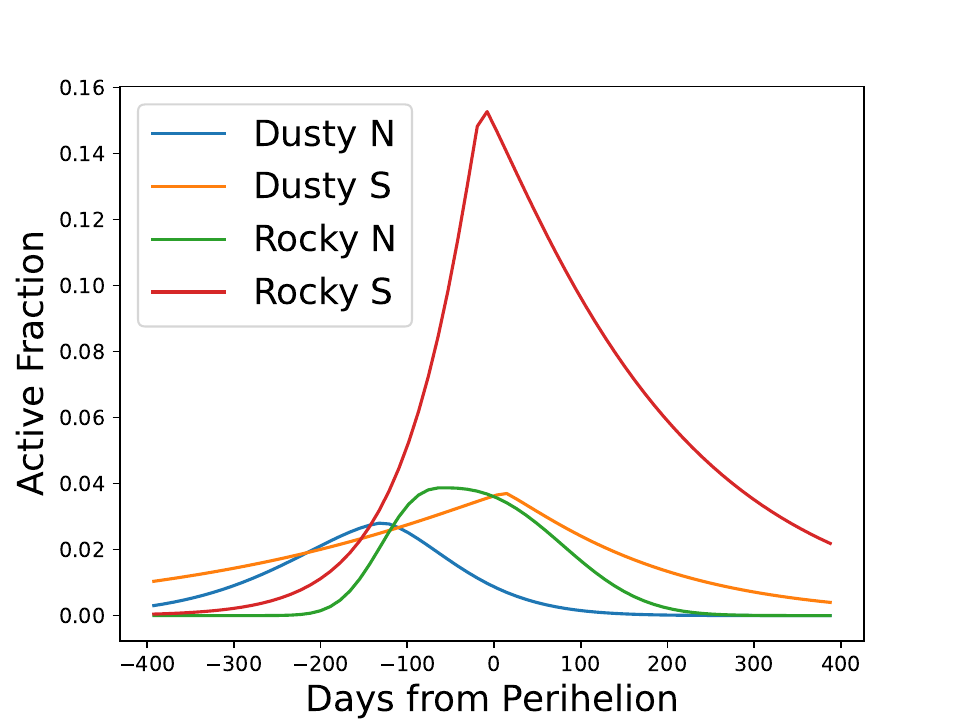}}%
\qquad
\subfloat{\includegraphics[scale=0.3]{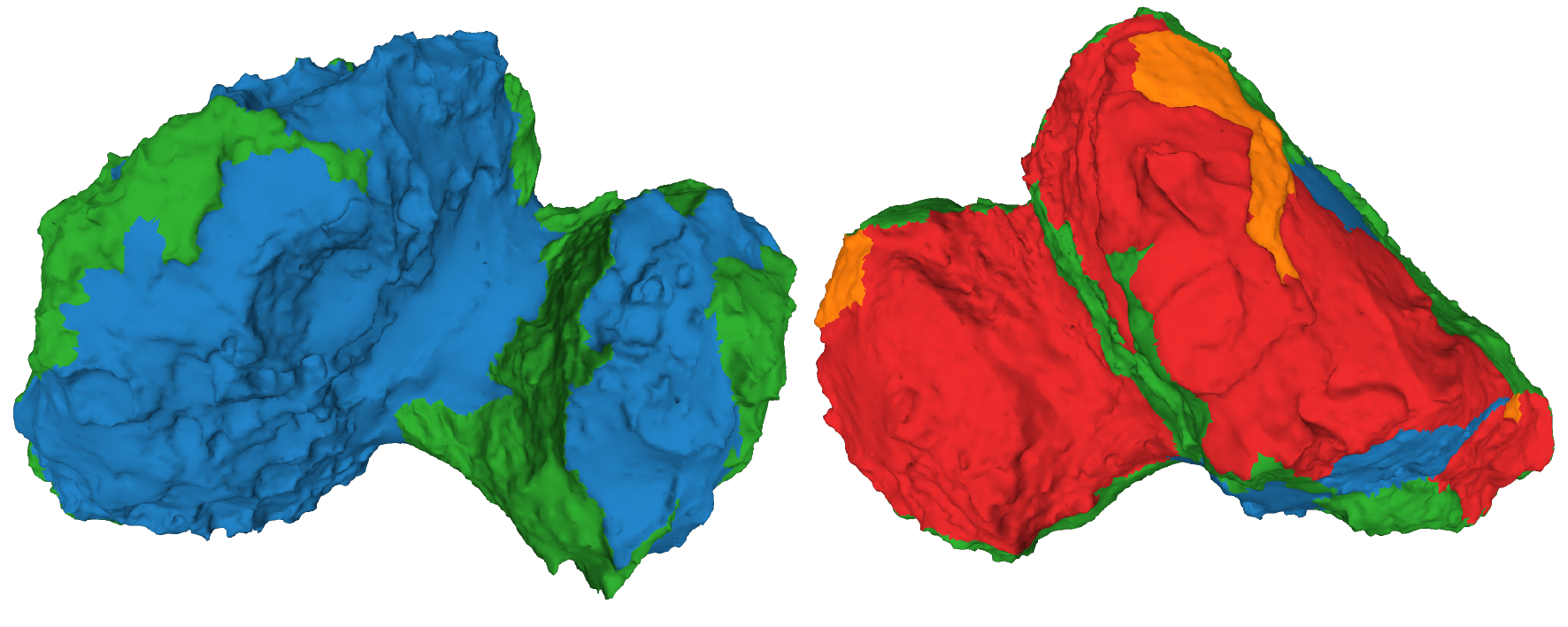}}%
\caption{Super-regions used in the best-fit solutions, shown with their EAF as a function of time for solution one and mapped onto the shape model with the corresponding colours.}
\label{Plot_map_run62}
\end{figure}

Based on the results of Sec.~\ref{results}, we decided to pursue a simple model, splitting terrains that cross both hemispheres into northern and southern areas. We first combined the \citet{Thomas2018} `Dust', `Brittle', `Smooth', and `Depression' terrains into a single `Dusty' super-region. This Dusty, and the remaining `Rocky', super-regions were then split by their NGA normal component at perihelion into northern and southern super-regions, resulting in a total of four super-regions. This splitting by an acceleration component, i.e.~facet orientation rather than geographic latitude, was motivated by the fact that it is surface orientation not latitude that actually controls insolation and outgassing from a particular facet. The resulting super-region definition is mapped in Figure \ref{Plot_map_run62}, and the model contains $4\times6=24$ free-parameters.

\begin{figure*}
\begin{tabular}[b]{@{}cc@{}} \\[-\dp\strutbox]
\subfloat{\includegraphics[width=6cm]{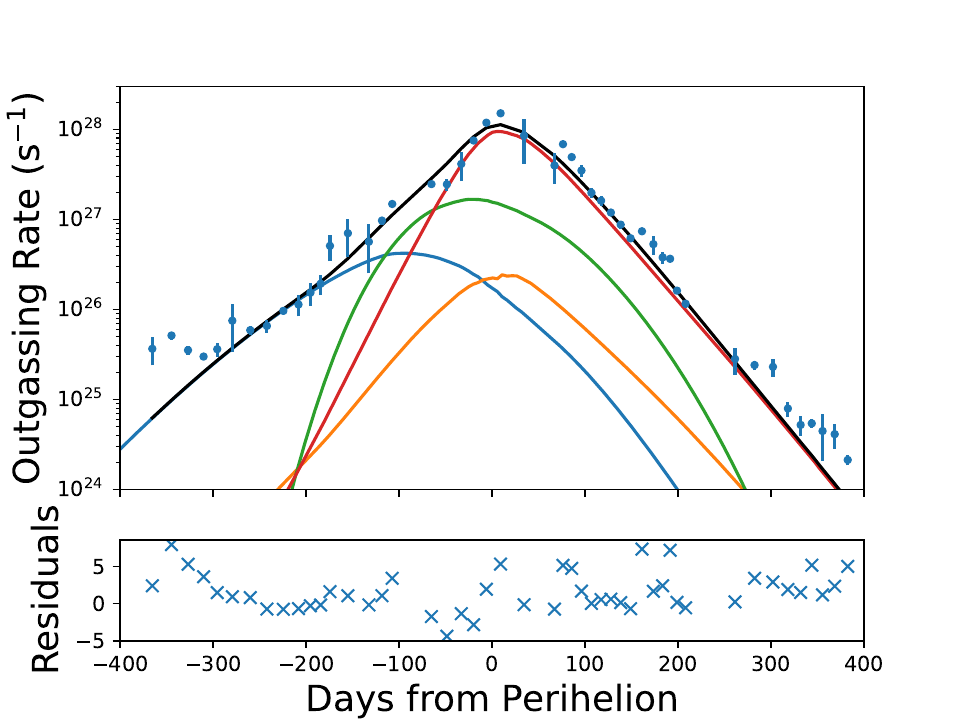}}
\subfloat{\includegraphics[width=6cm]{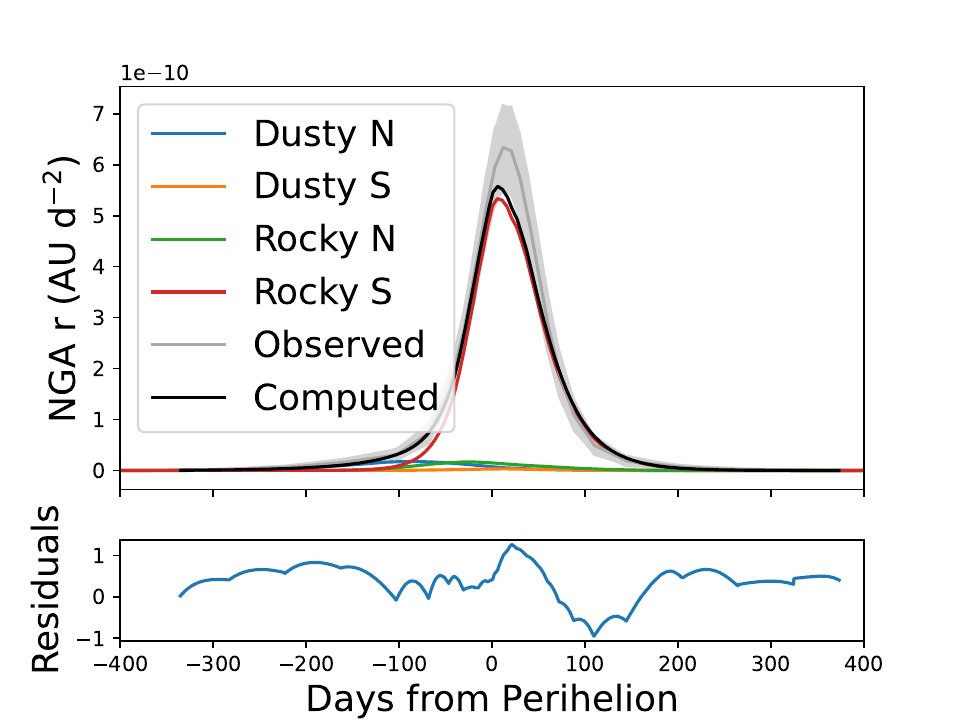}}
\\
\subfloat{\includegraphics[width=6cm]{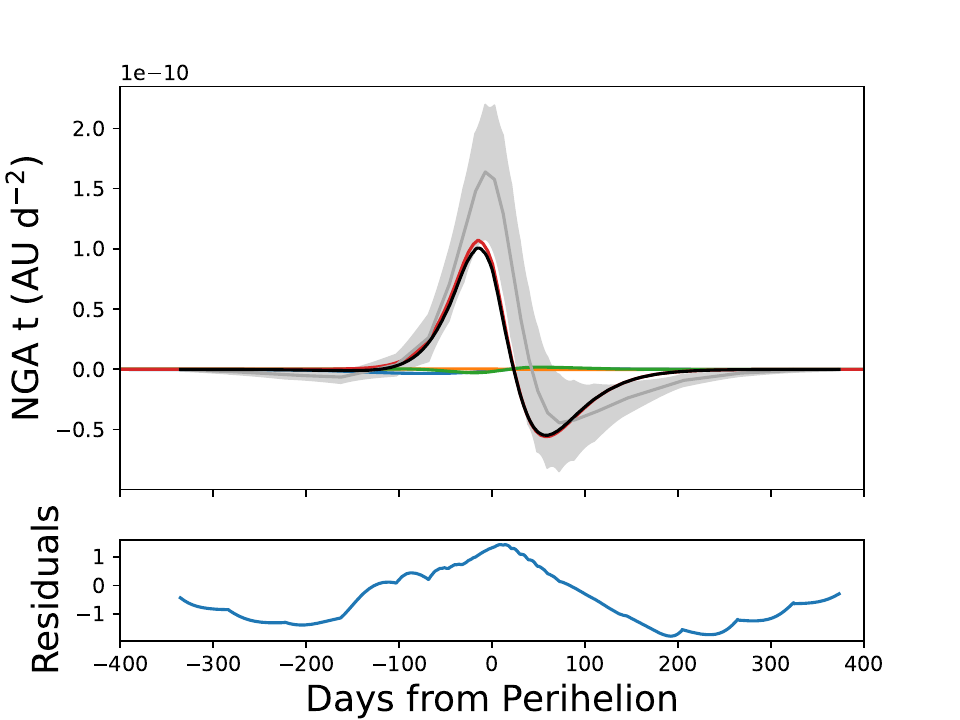}}
\subfloat{\includegraphics[width=6cm]{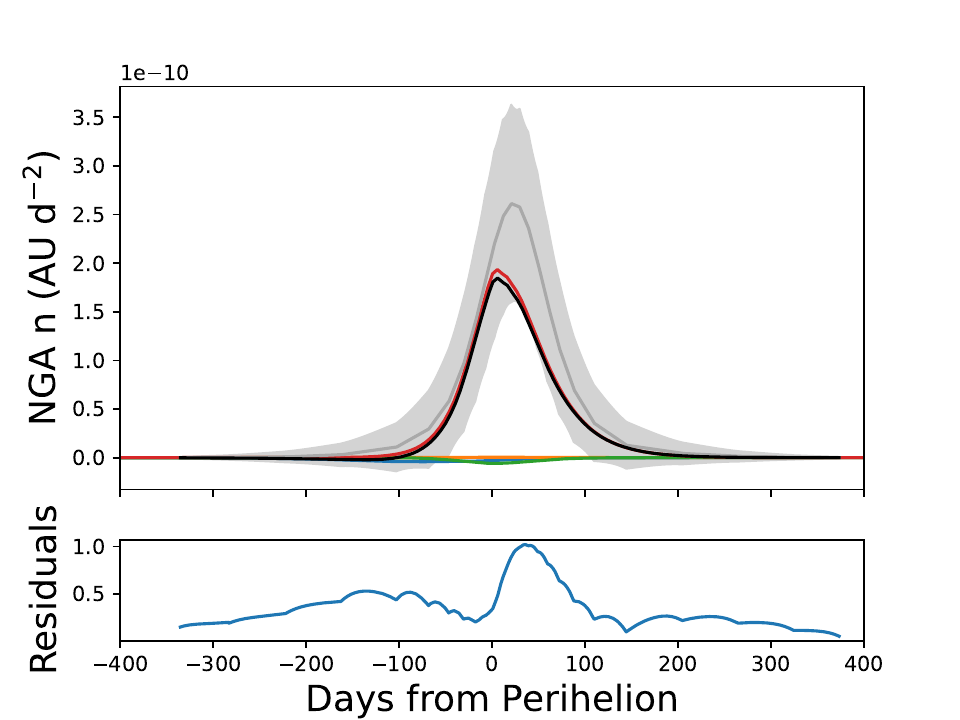}}
\\
\subfloat{\includegraphics[width=6cm]{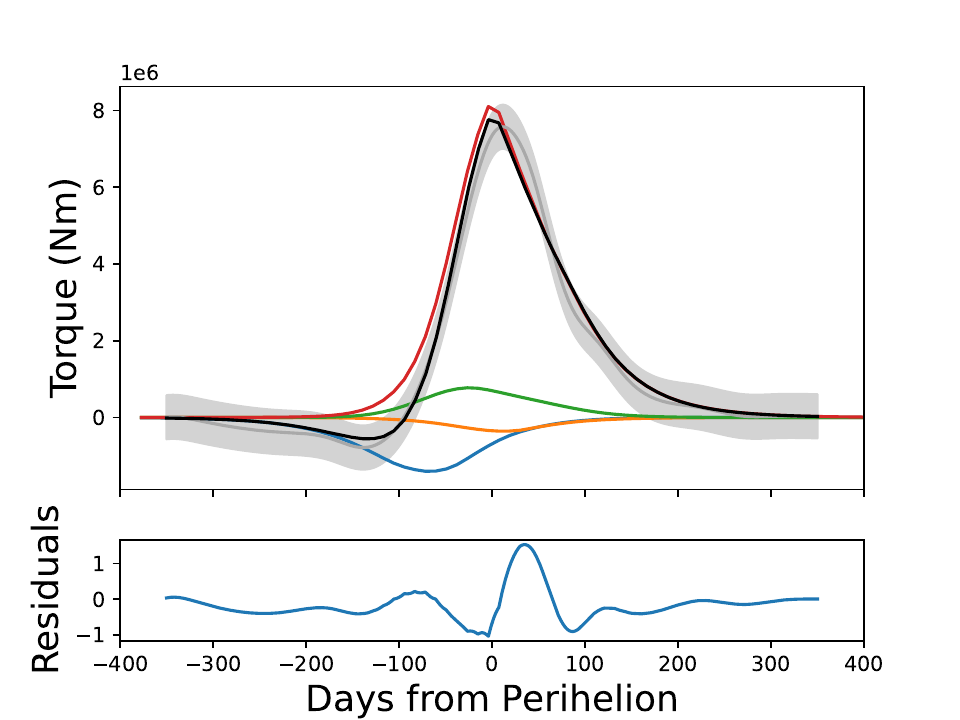}}
\subfloat{\includegraphics[width=6cm]{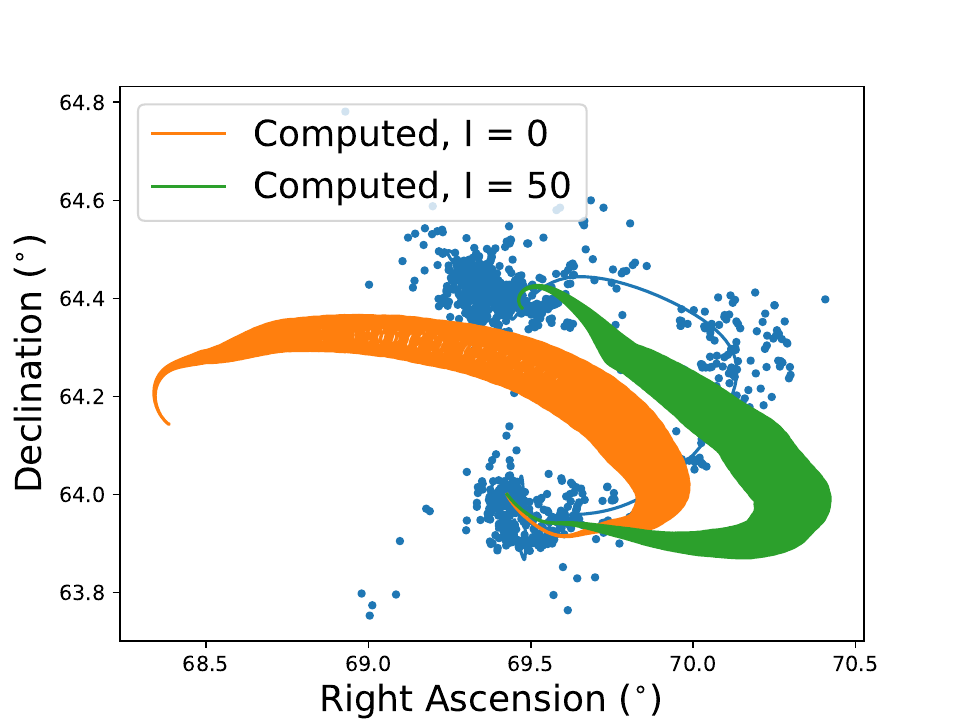}}
\end{tabular}\hfill
\begin{minipage}[b]{0.3\textwidth}
\caption{Results of the best-fit solution (solution one) and the normalised residuals for (from top left): total water production rate, radial NGA component, tangential NGA component, normal NGA component, z component of the torque, and pole-orientation. Individual contributions from each super-region are shown alongside the totals and the observational data (see text in this paper, \citealp{Attree2019}, and \citealp{Attree2023} for references).}
\label{Plot_results}
\end{minipage}
\end{figure*}

Fig.~\ref{Plot_results} shows the results of the fit against the observed total water-production rate, radial, tangential, and normal NGA components (from \citealp{Bologna}), and z component of the torque, as well as the match to the pole orientation. As a reminder, this last dataset is not included in the fitting routine but is output by the model for comparison. As can be seen, this model provides a significant improvement over the previous solutions, with an overall excellent match to all the fitted datasets, and a reasonable match to the pole orientation. The total rotation-period change of 22.5 minutes is also very close to the measured 21 minutes \citep{Mottola14, Jorda2016}. For the pole-orientation evolution, the raw data are plotted alongside the smoothed data used in the previous plots, demonstrating that, while our computed solution does display significant diurnal oscillations and is not a perfect match to the data, it does show roughly the correct magnitude of change and a similar scatter. Considering the minor rotation of the pole-orientation evolution achieved in Section \ref{check} above, we reran the fit using a thermal inertia of $I=50$ thermal inertia units on the 10k shape-model. Only very minor differences were observed in the NGA and water-production curves, but the pole-orientation evolution was indeed rotated slightly, as shown in the bottom right of Fig.~\ref{Plot_results}, now coming very close to the observations. The remaining differences between the modelled and observed pole-orientation evolutions may be explained by variations in this thermal inertia, particularly spatially. Minor albedo or emissivity variations may also play a role, as will the particularities of discretising the regions on the shape model and the initial rotation conditions.

From the water-production plot in Fig.~\ref{Plot_results}, it can be seen that northern dusty terrain dominates the total water outgassing until roughly the inbound equinox; while, after the equinox, rocky southern terrain takes over and dominates until the end of the Rosetta period. Around the equinox itself there is a brief period when rocky northern terrain provides the largest contribution. In terms of the NGAs and NGTs, rocky southern terrain completely dominates (in agreement with our previous work, as well as other studies: \citealp{Kramer2019}) while the addition of a relatively active northern dusty terrain produces the negative peak in torque seen before perihelion. Overall, southern dusty terrain contributes little to the production, NGA, or NGT, in line with its small geographical coverage. As shown in Figure \ref{Plot_map_run62}, the above results are achieved by a high activity peak of EAF $\approx15\%$ in the southern rocky terrain at perihelion, while all other terrains peak at around $~2-4\%$, with northern regions constrained to fall off in EAF before perihelion. This reduction in relative activity for the north as the comet approaches the Sun is a common feature of model fits here, and is important for simultaneously fitting the peak NGA, NGTs, and water-production.

\begin{table}
\caption{Momentum transfer coefficients, $\eta$, in the best-fit solutions.}
\begin{center}
\begin{tabular}{ccccc}
Solution & Dusty N & Dusty S & Rocky N & Rocky S  \\
\hline
One & 0.56 & 0.15 & 0.12 & 0.62 \\
Two & 0.85 & 0.10 & 0.10 & 0.57 \\
\end{tabular}
\label{table_results}
\end{center}
\end{table}

In terms of momentum transfer, Table \ref{table_results} shows that rocky southern terrain has the highest basic value of $\eta$, to go-along with its high EAF. Dusty northern terrain, meanwhile, has similar if slightly lower $\eta$; and the two other regions must be much less efficient at transferring momentum. It must be remembered, however, that this base value is multiplied by the temperature dependent factor in the brackets in Eqn.~\ref{eta} when calculating the force-per-facet, enhancing $\eta$ by a factor that peaks at perihelion and in the daytime, concentrating NGAs and NGTs diurnally and seasonally as shown at the top of Figure $\ref{Plot_insolation}$. At their most active times, when their EAF curves peak, the diurnally averaged 'total' $\eta$ values of northern dust and southern rock are around 0.75 and 1, respectively, with the other two regions still much lower.

Considering the residuals to the water-production curve (top left in Fig.~\ref{Plot_results}), we see that there are two outlier points located near perihelion and the first equinox, respectively. These points' relatively small error-bars combined with their high values drive the overall fit statistic, limiting the ability of the fitting routine to improve the model. The error-bars take into account only the limited surface coverage of the ROSINA (Rosetta Orbiter Spectrometer for Ion and Neutral Analysis) instrument \citep{Laeuter2020}, and therefore may underestimate the actual uncertainty, which is visible in the scatter of the other data-points. We therefore reran the fit with these two production points removed, in order to evaluate their effect.

\begin{figure}
\includegraphics[scale=0.5]{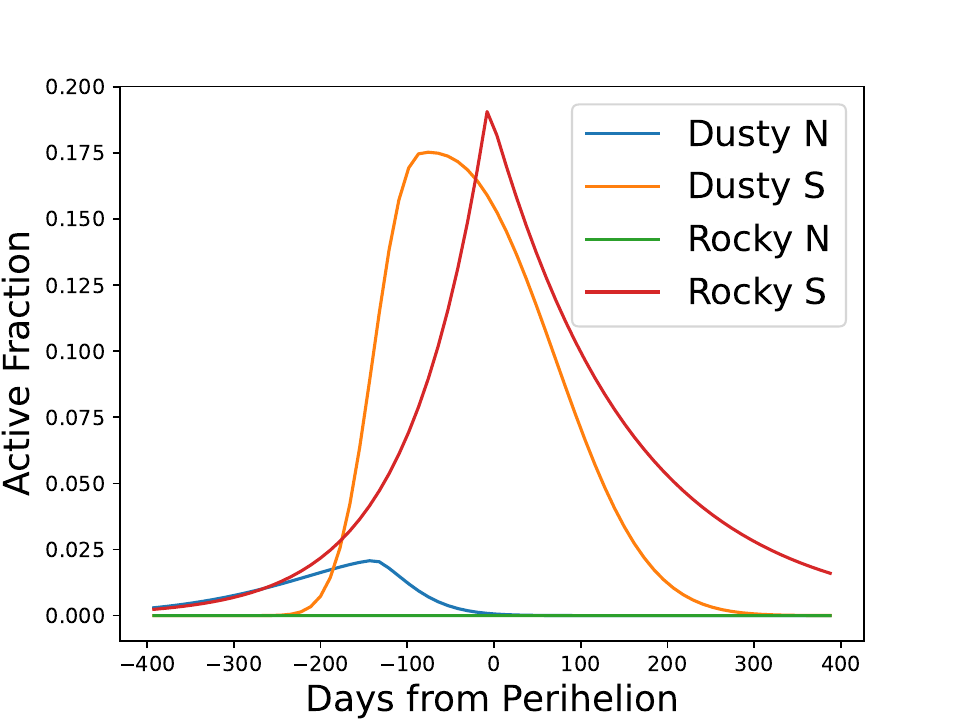}
\caption{EAF as a function of time in best-fit solution two.}
\label{Plot_map_run62_other}
\end{figure}

\begin{figure*}
\begin{tabular}[b]{@{}cc@{}} \\[-\dp\strutbox]
\subfloat{\includegraphics[width=6cm]{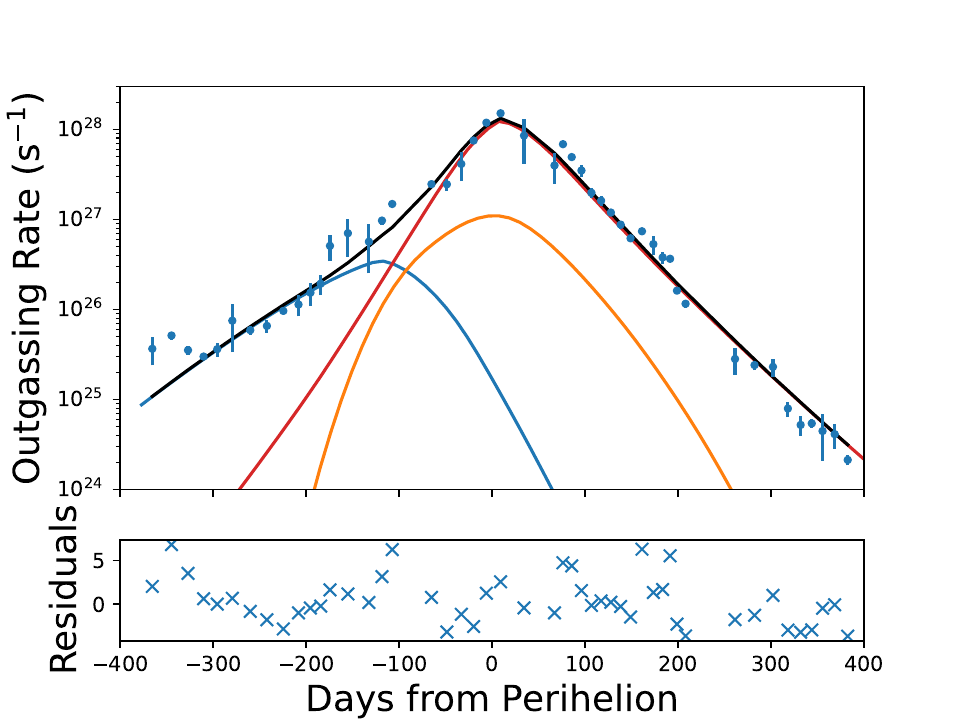}}
\subfloat{\includegraphics[width=6cm]{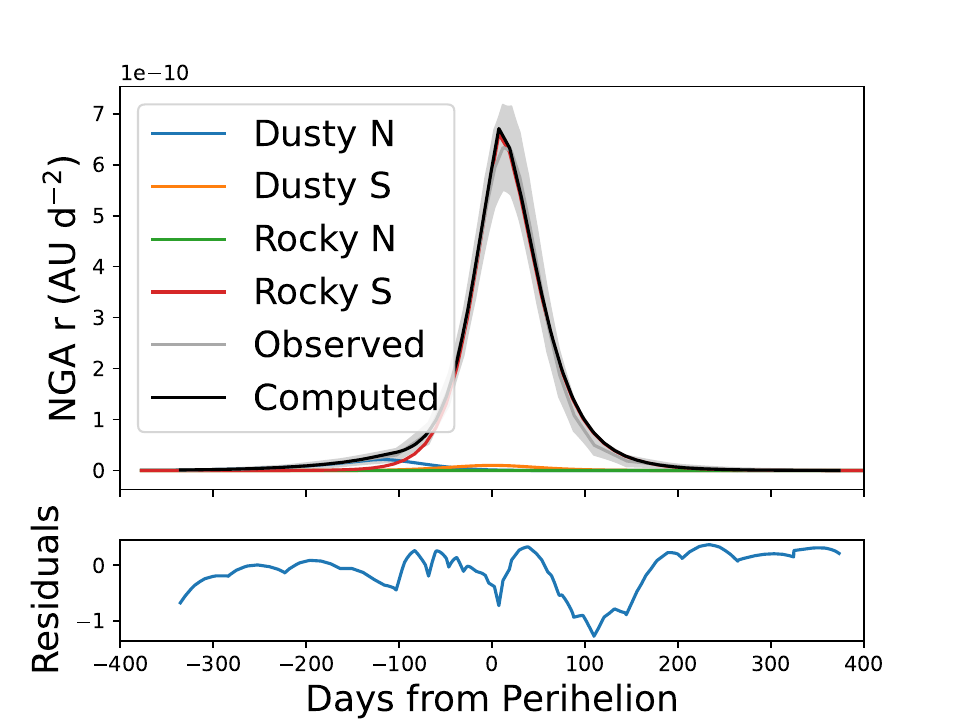}}
\\
\subfloat{\includegraphics[width=6cm]{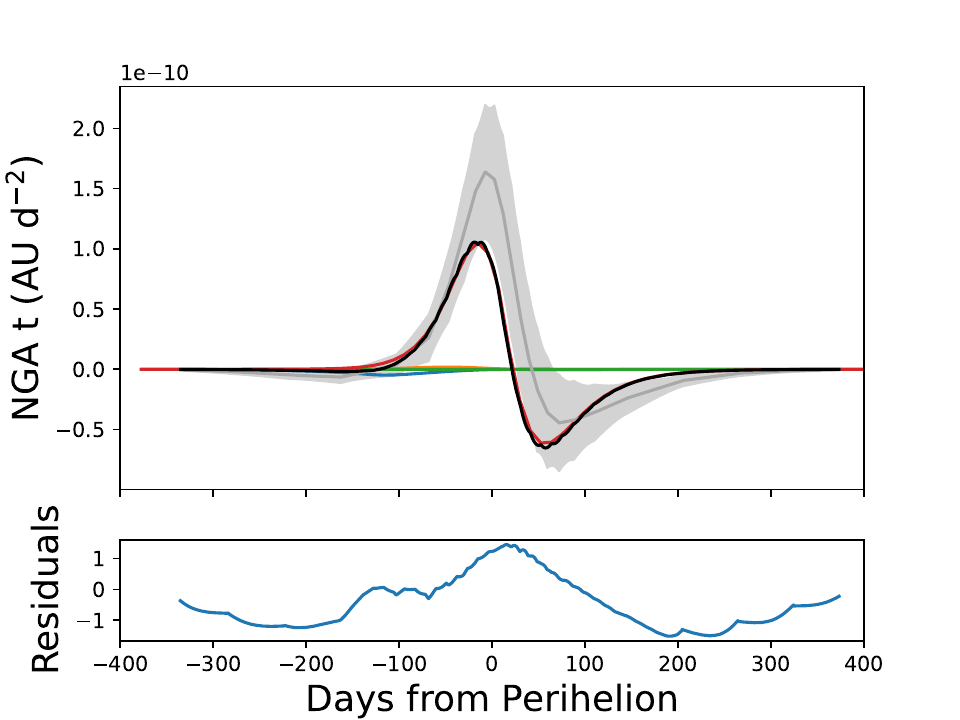}}
\subfloat{\includegraphics[width=6cm]{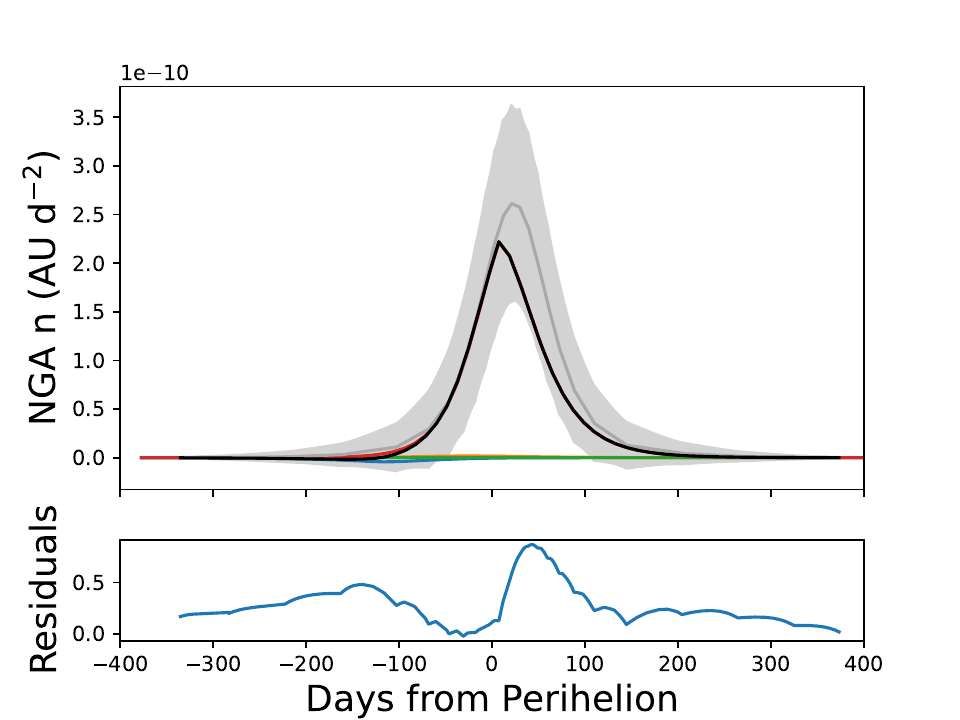}}
\\
\subfloat{\includegraphics[width=6cm]{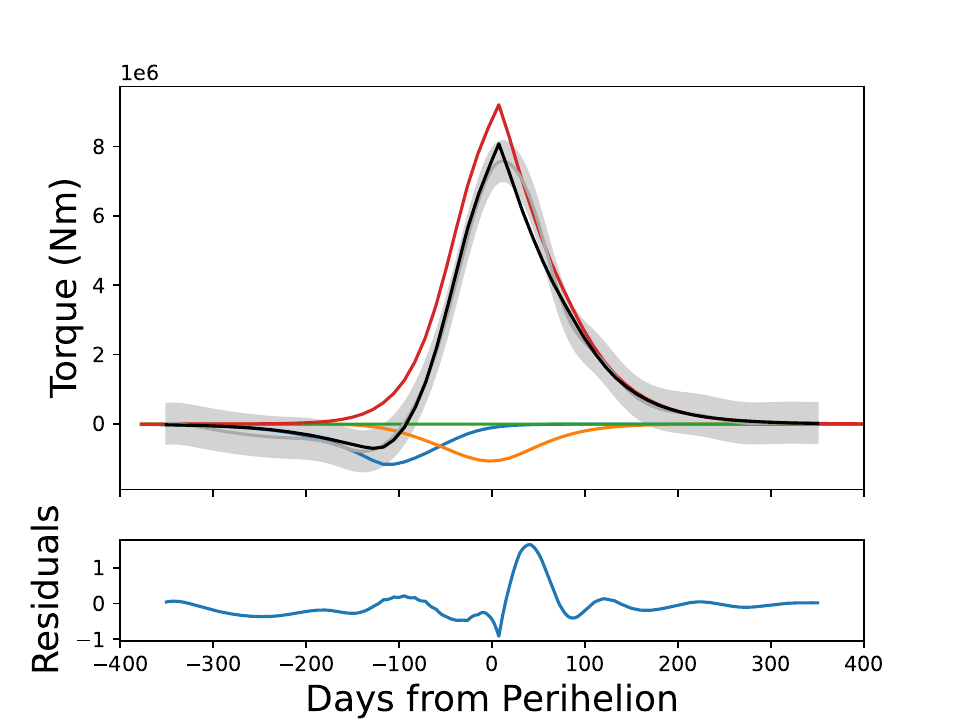}}
\subfloat{\includegraphics[width=6cm]{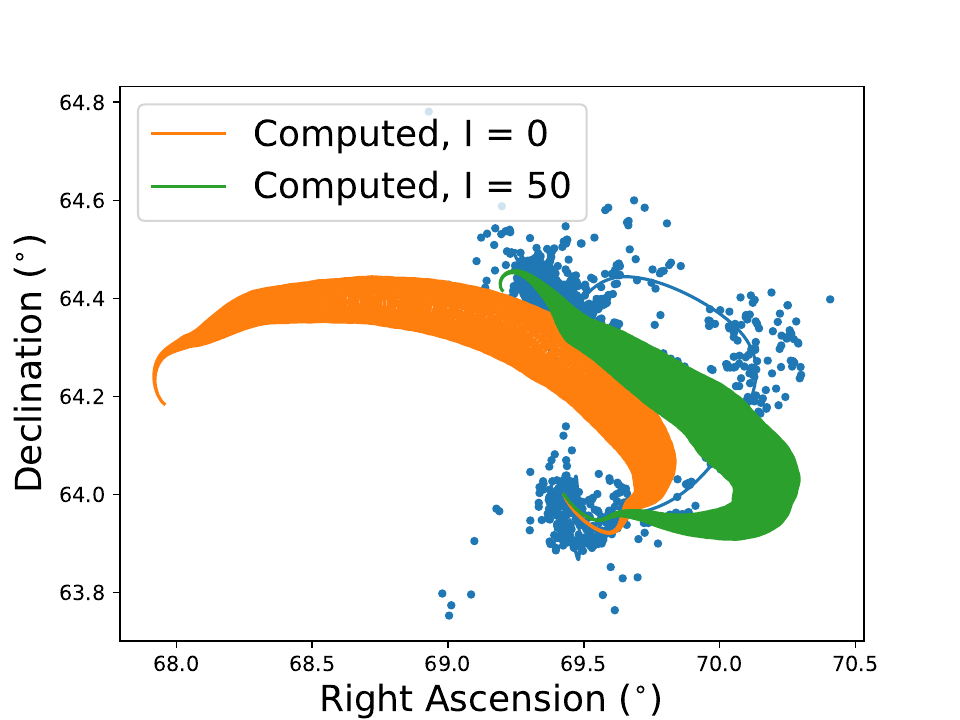}}
\end{tabular}\hfill
\begin{minipage}[b]{0.3\textwidth}
\caption{Results of the best-fit solution (solution two) and the normalised residuals for (from top left): total water production rate, radial NGA component, tangential NGA component, normal NGA component, z component of the torque, and pole-orientation. Individual contributions from each super-region are shown alongside the totals and the observational data (see text in this paper, \citealp{Attree2019}, and \citealp{Attree2023} for references).}
\label{Plot_results_other}
\end{minipage}
\end{figure*}

Figures \ref{Plot_map_run62_other} and \ref{Plot_results_other} show the EAF curves and fit to the data for this new solution, which we refer to as solution two. The match to the NGAs and NGTs is very similar in this case, while the production residuals are improved, leading to an overall somewhat improved fit statistic. The rotation-period change is now even closer to the measurements at 21.4 minutes. From Fig.~\ref{Plot_map_run62_other}, we see that solution two has a slightly higher southern rocky activity ($\approx18\%$), and a much higher southern dusty activity than before, alongside zero EAF in the rocky northern terrain. Northern dust falls off as before. The total water production is then dominated by the northern dust, smoothly transitioning to southern rock at the equinox, with a minor contribution from southern dust, again due to its small area. Table \ref{table_results} shows that the momentum transfer coefficients follow a similar pattern to before, but now with northern dusty terrain having the highest values, over the southern rock. When multiplied by the enhancement factor of Eqn.~\ref{eta}, the total momentum transfer coefficients are around 1.1 and 0.9 for these two regions, respectively. Rocky northern terrain's momentum transfer coefficient has little meaning due to its negligible EAF. Similarly to above, rerunning solution two with thermal inertia led to almost exactly the same NGA curves and z component of the torque, while the pole-orientation evolution was slightly rotated (bottom right of Fig.~\ref{Plot_results_other}).

Minor variations in these fits could be achieved by varying the weighting and error treatment of the various datasets, but the competing restrictions placed by the NGA and NGT curves ensured that no significant improvements could be found using this model. We thus, finally, consider this our best-fit model for 67P's water outgassing pattern; with solution two our preferred choice, due to its improved fit statistic.

\section{Implications for cometary activity models}
\label{discussion}

In general, equinox represents a distinct change in the activity, from being driven by the northern to southern hemispheres, resulting in different behaviour (as has also been noted in other works, e.g.~\citealp{Davidsson2022}). Regarding NGAs, the accelerations are small in magnitude outside of the equinoxes and therefore not useful in constraining activity at large heliocentric distances, but are completely dominated by activity between the two equinoxes. This means that they are most sensitive to the south (which dominates activity between the equinoxes), but the high normal and tangential NGA components do constrain northern activity during this period to be low, while simultaneously constraining the southern activity to be relatively very high and rather efficient at transferring momentum.

Conversely, NGTs are sensitive to both northern and southern activity during the Rosetta mission proximity operations, e.g.~the first, negative, peak in the z torque component (corresponding to an observed increase in the spin-period) is driven by activity from the north at the first equinox, whereas the second, positive, peak (corresponding to the overall, larger decrease in spin-period) is driven by southern activity (specifically, the northern dust and the southern rock, respectively, in our best-fit model). Likewise, the pole-orientation evolution appears sensitive to all areas of the cometary surface.

Finally, the water-production rate curve of \citet{Laeuter2020} requires high relative activity at perihelion (although peak production is slightly less than estimated by \citealp{Hansen}; see discussion in \citealp{Laeuter2020}), but also an asymmetry, with higher production for a particular time pre-perihelion as for the corresponding post-perihelion time.

From the above, and the previous three sections, a number of conclusions regarding cometary activity on 67P can be made.

\begin{figure}
\subfloat{\includegraphics[scale=0.5]{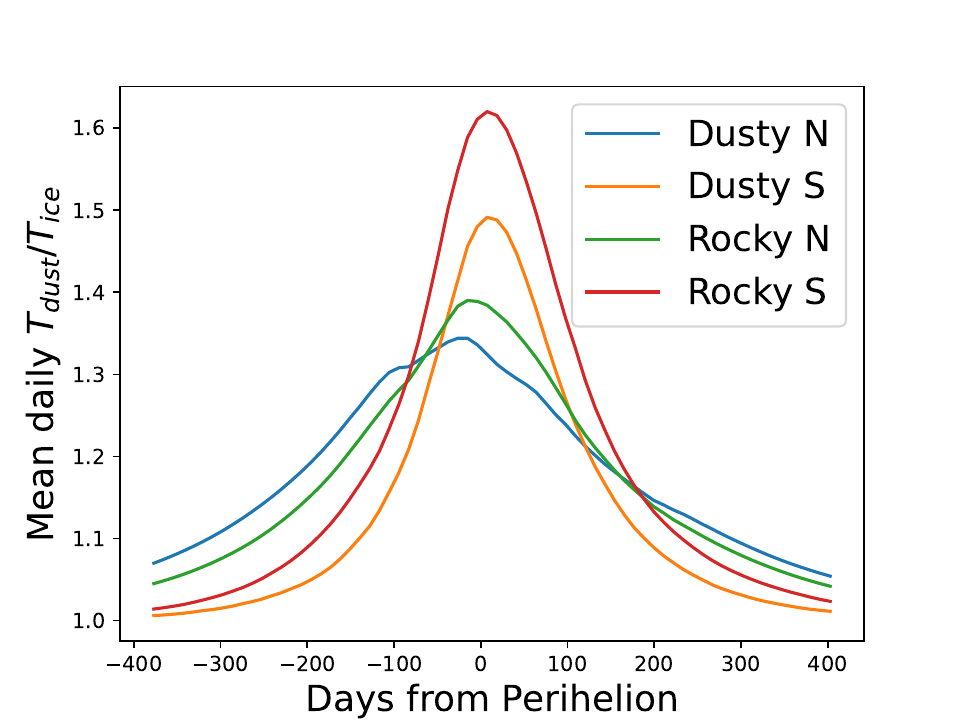}}%
\qquad
\subfloat{\includegraphics[scale=0.5]{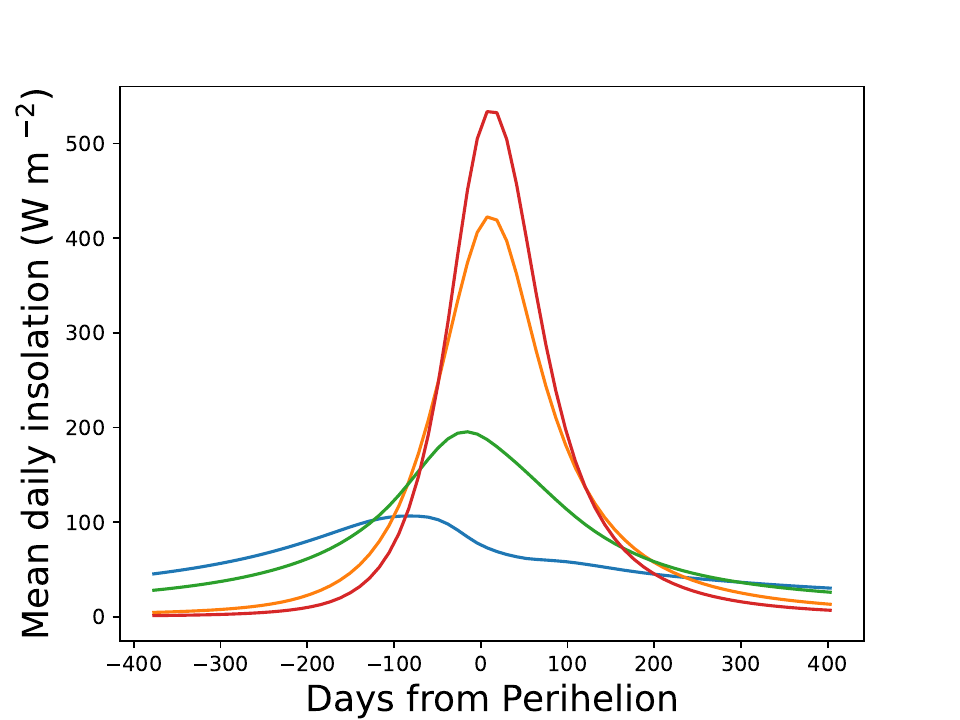}}%

\caption{Top: mean daily momentum enhancement factor ($T_{dust}/T_{ice}$, as used in Eqn.~\ref{eta}), and bottom: mean daily insolation, averaged over each of the four super-regions used in our best-fit solutions.}
\label{Plot_insolation}
\end{figure}

\begin{enumerate}
    \item Water sublimating from the nucleus is the primary driver of non-gravitational forces and torques. The analysis of the magnitudes of the NGA and water production curves in Section \ref{method} is consistent with water sublimating from the surface, rather than other gas species or water from extended sources, being the main driver of 67P's non-gravitational dynamics.
    \item Thermal inertia and self-heating have only minor effects on non-gravitational forces and torques. Thermal inertia has, in some previous studies, been considered as driving the non-radial components of the NGA but, at least for our best-fitting model of 67P, the effects of a moderate thermal inertia on NGA and NGT are limited to a minor rotation of the pole-orientation evolution.
    \item Spatially uniform activity cannot explain 67P's non-gravitational dynamics. The analysis from Section \ref{check}, together with previous works, shows that when the response of the surface to insolation is the same everywhere on the comet then the resulting pole-orientation evolution does not match the Rosetta observations. This is the case even when the insolation response has a non-linear dependence. Areas of the comet with otherwise similar instantaneous illumination levels (see \citealp{Attree2023}, Fig.~12 and Fig.~\ref{Plot_insolation}, below) must supply different amounts of outgassing in order to generate the observed NGTs. Note that this does not completely rule out models such as that of \citet{Fulle2020}, but imposes strong constraints on them in terms of the surface response to solar energy input: some areas must show an enhanced outgassing compared to others, by material differences or by a feedback effect from long-term insolation difference or fall-back, for example.
    \item Spatially uniform momentum transfer cannot explain 67P's non-gravitational dynamics. Likewise, it appears impossible to simultaneously fit the non-gravitational forces and torques using models with a single, fixed value of $\eta$, the momentum transfer coefficient. Instead, $\eta$ must vary across the surface, and/or with insolation, with different terrain types being more or less efficient at collimating the outgassing, and a strong enhancement with insolation being favoured.
    \item Different terrain types have different instantaneous responses to insolation. Improvements to the fit of \citet{Attree2023} are achieved by having the south-facing rocky terrain exhibit a different activity response to that which faces north. Only in this way can the peak NGAs at perihelion be fit. This is the case whether splitting the surface by broad geographic region, as in \citet{Thomas2018}, or by feature-scale geomorphology, as in \citet{Birch17}. In both cases, terrain which appears visually the same behaves differently depending on whether it faces generally north or south.
\end{enumerate}

The second and last points can be more easily visualised when plotting the averaged daily insolation curves for each super-region, as done in Figure \ref{Plot_insolation}. Southern-facing terrains clearly receive the most insolation, but the location of some 'northern' rocky terrains means that they too receive a large energy input at perihelion. Insolation to the northern dust peaks around the first equinox and then begins to decline to perihelion. When compared to our best-fitting EAF curves (Figs.~\ref{Plot_map_run62} and \ref{Plot_map_run62_other}), there is a general correlation between increasing insolation and peak EAF (i.e.~blue then orange then red curves, in order of increasing both), but the green, rocky north curve does not follow the trend. Likewise, the shapes of the two dusty curves are different from the rocky ones in solution two, and both solutions have peak EAF in the dusty north reached before peak insolation, followed by a decline whilst insolation remains high.
A similar but inverse decoupling of insolation and the gas emission rate was observed by \citet{Laeuter2022} for CO$_2$. In their northern patch regions close to the perihelion passage, the diurnally averaged insolation decreases and the gas production increases at the same time.

Overall then, what does this imply for the surface material of 67P? Our best-fit model reinforces the idea of southern rocky terrain as being composed of relatively pristine, volatile-rich material (e.g.~see \citealp{Cambianica2020, Herny2021, Davidsson2022}) that produces a high water outgassing rate with a very steep increasing response to insolation. This may be similar to the Water Enhanced Blocks (WEBs) composed of pebbles as proposed by \citet{Fulle2019, Fulle2020}; and \citet{Ciarniello2021, Ciarniello2023}, or may reflect an enhancement in outgassing from a more deeply penetrating heat-wave due to the constant illumination of polar day \citep{Skorov2020}. This latter suggestion is tentatively supported by the fact that Southern rocky terrain must also have a relatively large momentum transfer coefficient (especially in solution one), i.e.~a sublimation that is very collimated, possibly because if happens at a deep level or is otherwise efficiently channelled. In our preferred solution two, however, northern dust has an even higher momentum transfer, complicating this interpretation. The general improvement to the fits using an insolation dependent momentum transfer, as described by Eqn.~\ref{eta}, does imply a correlation between increasing solar input and increasing gas temperature which is greater than the simple energy balance model. Such an enhancement is less pronounced in the \citet{Fulle2020} model due to its lower surface temperatures compared to a pure dust surface. Surfaces that are smooth on a local scale might also be expected to have a higher momentum transfer overall than rough surfaces, where parts of the gas flow will `cancel out'. This would suggest higher $\eta$ in the smooth, dusty plains, which is seen in solution two but not in solution one. On the other hand, rough surfaces may be hotter than smooth ones, depending on the scale of the roughness. Outgassing speed, mass and momentum flux are clearly coupled to the material structure and surface shape in a complex way that requires further thermophysical modelling to disentangle.

North-facing rocky terrains, although similar in visual appearance, behave differently to south-facing ones, with much lower peak outgassing and momentum transfer. In our solution one, peak EAF for north-facing rock is similar to that of south-facing dust, whereas solution two prefers no outgassing at all from the former region. Both solutions agree, however, that this terrain produces less outgassing than south-facing rock. This may reflect differing thermal histories in near-vertical cliffs versus flat plains in the northern and southern hemispheres, respectively, or it may represent a structural or compositional difference. In the latter case, cliffs such as those in Geb would have different properties to nearby areas on both lobes such as Neith, Wosret, and Bes, while parts of Anhur differed from one-another. It is difficult to further constrain these differences with the available data, however, while outgassing and other Rosetta measurements generally do not suggest such sharp compositional distinctions (see, e.g.~\citealp{Laeuter2020, Patzold16, Groussin2019}).

Dusty material has a differently shaped outgassing curve; in particular, north-facing dust increases its EAF with increasing insolation, but reaches peak values much less than the southern rocky terrain, and then begins to decline whilst insolation remains high. The northern plains' contribution to the overall water production then falls rapidly and becomes negligible towards perihelion, providing even less contribution than northern polar winter would suggest, while it also does not recover after the second equinox, leading to the asymmetric production curve mentioned above. This may be because northern dusty plains are composed of fall-back material that has been depleted in volatiles \citep{Cambianica2020} and it begins to 'run out' of ice with increasing insolation. Alternatively, ongoing fall-back of material from the more active south may be burying the plains in an ever deeper dry-dust layer as its activity decreases. The fact that 67P's activity appears relatively consistent between orbits, both in terms of its outgassing curve \citep{Snodgrass2017} and rotation-rate changes \citep{Mottola2014, Jorda2016}, means that, either way, the northern plains must be resupplied by fresh fall-back with at least some volatile content in order to repeat the cycle \citep{Keller2017}. Conversely, it must not be reactivated after perihelion in this same orbit despite the fact that insolation conditions are similar to the inbound orbit (Fig.~\ref{Plot_insolation}). \citet{Davidsson2021} and \citet{Davidsson2022} together suggest a possible explanation whereby $\sim$ cm-sized fall-back particles develop of an insulating mantle during their flight-time, allowing them to maintain  $85-95\%$ of their overall water ice content. Then, either the slightly differing insolation conditions combined with the mantle, restricts their outgassing until a more favourable insolation inbound next orbit favours gas release, or they suffer substantial alteration during aphelion due to thermal fatigue processes. This alteration of the characteristics could facilitate the release of volatile material on the next inbound orbit. \citet{Keller2017} instead explain the reactivation as being limited to Hapi, which remains shadowed and cold even after the rest of the dusty north begins to be illuminated again after perihelion, allowing volatile-rich fall-back particles to survive here until the next apparition. The timing and magnitude of the decreasing outgassing response should be studied in more detail to constrain these scenarios.

South-facing dusty terrain covers a very small area, and hence it is hard to constrain its activity. Our preferred solution (solution two), however, has it behaving a similar way to north-facing dust, but with the more intense insolation leading to a higher peak EAF. Once again though, EAF falls off before receiving peak insolation, which may reflect the same depletion of volatiles in the partially dehydrated fall-back material as above. Once again, this material must be reactivated on the next perihelion passage to maintain 67P's repeating activity. South-facing dusty areas would then represent the same kind of material as north-facing ones, just subjected to a higher peak insolation.

The above results would represent a modification, rather than an outright refutal of the model of \citet{Ciarniello2023}, which posits a perihelion activity following the \citet{Fulle2020} outgassing curve but controlled by the ejection of large chunks by CO$_{2}$. Here, we need an even greater concentration of outgassing and an enhancement in the momentum transfer coefficient in particular areas of the south-facing consolidated terrain, combined with a decreasing northern activity, and a low relative activity and momentum transfer in other southern areas, like the dusty ones, and north-facing cliffs. As noted above, this could come from a combination of volatile depleted fall-back in some areas, and an enhancement in other areas that could indeed be related to CO$_{2}$ emission (which peaks in the south: \citealp{Laeuter2020}), or to polar day. Our best-fit model also has high activity continuing in the southern rock for longer than the \citet{Fulle2020} model would suggest. This may come from reactivated fall-back material as suggested by \citet{Ciarniello2023}, although this would have to be confined to only our southern rocky region, or it may be related to continued  CO$_{2}$ driven erosion with a thermal lag-time. CO$_{2}$ appears to have little direct effect on the NGAs and NGTs, but its interaction with the water-production rate, via erosion and the exposure of fresh material, clearly needs further investigation. This requires full, time-dependent thermophysical modelling of the activity, including CO$_{2}$ sublimation and chunk ejection, which is under development and will be the subject of a future paper. Finally, \citet{Ciarniello2023} also posit water emission from extended sources in the coma before the first equinox, which is difficult to constrain with NGAs and NGTs as mentioned above.

\section{Conclusion}
\label{conclusion}

In this paper we have extended the previous work on modelling the non-gravitational trajectory and rotation of 67P/Churyumov-Gerasimenko using a thermal activity model. In particular, we have compared the various NGA extractions available in the literature, before formally investigating the effects on the modelling of shape-model resolution, self-heating, thermal inertia, a more detailed surface mapping, and varying the momentum transfer coefficient. This last parameter was found to be particularly important, and by a more complex treatment of it, we are much better able to match the observational data of Rosetta.

A number of conclusions can be drawn from this modelling effort:
\begin{itemize}
\item Water sublimating from the nucleus is the primary driver of non-gravitational forces and torques.
\item Thermal inertia and self-heating have only minor effects on non-gravitational forces and torques.
\item Spatially uniform activity cannot explain 67P's non-gravitational dynamics.
\item Spatially uniform momentum transfer cannot explain 67P's non-gravitational dynamics.
\item Different terrain types have different instantaneous responses to insolation.
\end{itemize}

Further, our best-fit solution supports a model of 67P's surface being mostly comprised of pristine, volatile-rich south-facing consolidated material with a high outgassing flux, steep response to insolation, and large gas momentum transfer coefficient. Dusty north-facing terrain, on the other hand, either depletes its volatile component or is buried in fall-back as the comet approaches the Sun. The small area of south-facing dusty terrain must also have a low-to-intermediate outgassing rate and momentum transfer coefficient and deplete its volatiles. Meanwhile, consolidated terrains that face north behaves differently to their southern counterparts, producing low-to-no water outgassing, and with a lower momentum transfer coefficient. In general, our modelled momentum transfer coefficient appears correlated with insolation, likely due to a strong enhancement in the gas temperature compared to a surface energy-balance model, as the dust it flows through is heated.

The above conclusions represent an empirical model of the outgassing response of different materials on the cometary surface that more complicated thermophysical models should seek to reproduce. It is unlikely that a single thermophysical model can reproduce all of these features simultaneously. Therefore our work supports the conclusion of spatially and/or temporally varying material properties over 67P's surface and orbit. It should be noted, however, that this observed difference may not necessarily stem from deep material or structural differences between different parts of the comet, but may arise naturally from the material's medium- to long-term response to the complicated insolation pattern.

Ultimately, 67P/Churyumov-Gerasimenko's non-gravitational orbit and spin place important constraints, alongside its water-production rate, on the distribution of outgassing and the activity mechanism in cometary material, contributing to our knowledge of the pristine nature of comets.

\begin{acknowledgements}
We thank the reviewer for their comments which improved the manuscript. We also thank Sam Birch and Abhinav Jindal for useful discussions and input, and Davide Farnocchia for kindly providing the NGA curves from the rotational jet model. N.A.’s contributions were made in the framework of a project funded by the European Union’s Horizon 2020 research and innovation programme under grant agreement No 757390 CAstRA. N.A. and P.G. acknowledge financial support from project PID2021-126365NB-C21 (MCI/AEI/FEDER, UE) and from the Severo Ochoa grant CEX2021-001131-S funded by MCI/AEI/10.13039/501100011033. This research was supported by the International Space Science Institute (ISSI) in Bern, through ISSI International Team project \#547 (Understanding the Activity of Comets Through 67P's Dynamics). The contribution of O.G. to this project was funded by the Centre National d'Etudes Spatiales (CNES).
\end{acknowledgements}

\bibliographystyle{aa}
\bibliography{Bibliography}

\begin{appendix}

\end{appendix}

\end{document}